# AN END-TO-END GSM/SMS ENCRYPTED APPROACH FOR SMARTPHONE EMPLOYING ADVANCED ENCRYPTION STANDARD (AES)

Wasim Abbas[1], Salaki Reynaldo Joshua[2], Asim Abbas[3] and Je-Hoon Lee[*4]

[1, 2, 4] Department of Electronics, Information and Communication Engineering, Kangwon National University, Samcheok-si, Republic of Korea
wasimabbas1603@kangwon.ac.kr, joshua@kangwon.ac.kr

[3]Division of Computer Science, Mathematics and Science, St. John's University, Queens NY 11439, USA
abbasa@stjohns.edu

[5]Department of Liberal Studies, Kangwon National University, Samcheok-si, Republic of Korea
jehoon.lee@kangwon.ac.kr

## ABSTRACT

*Encryption is crucial for securing sensitive data during transmission over networks. Various encryption techniques exist, such as AES, DES, and RC4, with AES being the most renowned algorithm. We proposed methodology that enables users to encrypt text messages for secure transmission over cellular networks. This approach utilizes the AES algorithm following the proposed protocols for encryption and decryption, ensuring fast and reliable data protection. This approach ensures secure text encryption and enables users to enter messages that are encrypted using a key at the sender's end and decrypted at the recipient's end, which is compatible with any Android device. SMS are encrypted with the AES algorithm, making them resistant to brute-force attempts. As SMS has become a popular form of communication, protecting personal data, email alerts, banking details, and transactions information. It addresses security concerns by encrypting messages using AES and cryptographic techniques, providing an effective solution for protecting sensitive data during SMS exchanges.*

## 1. INTRODUCTION

Short Message Service (SMS) enables wireless text messaging, restricted to 160 characters without graphics or images [1]. SMS messages are sent via Global System for Mobile Communication (GSM) [2]. Upon sending, the SMS messages are stored and forwarded by the Short Message Service Center (SMSC) to the intended mobile device [3]. SMSC utilizes the store-and-forward method, storing messages before forwarding them to the target device. If the target device's Home Location Register (HLR) is active, SMSC delivers the SMS message [4]. Confirmation of successful delivery is received by SMSC. Unencrypted SMS messages stored in SMSC are vulnerable to viewing and alteration by SMSC staff [5]. Some SMSCs retain SMS copies for profit and commercial purposes, accessing information from banks and other sources [6]. SMS is widely used for personal messages, information services, school alerts, bank transactions, delivery updates, notifications, and stock alerts [7].

Information security safeguards transmitted data against unauthorized access, disclosure, modification, or inspection [8]. Cryptography ensures data security by encoding and decoding data during transmission [10, 11]. To ensure SMS privacy, various cryptographic techniques are employed [12]. Secure end-to-end SMS encryption is essential for establishing a secure communication channel. DES and AES are commonly used cryptographic algorithms [13]. DES, with its 56-bit key, is vulnerable to brute-force attacks, while AES, employing larger keys (as shown in Table 1), and is resistant to such attacks [14].

The primary goal of this app is to study and become acquainted with the fundamental cryptographic algorithms. We are implementing SMS encryption through an Android application for mobile communication by using one of these algorithms to provide a secure environment for sensitive data while transmitting messages.

As a result, the following are the paper objectives:
- To provide adequate SMS security by utilizing encryption and decryption methods with a suitable algorithm to prevent fraud and theft of secret information when communicating private and sensitive data.
- We are working on creating a user-friendly system that is both simple to use and secure.
- To ensure that the system functions properly and without errors.

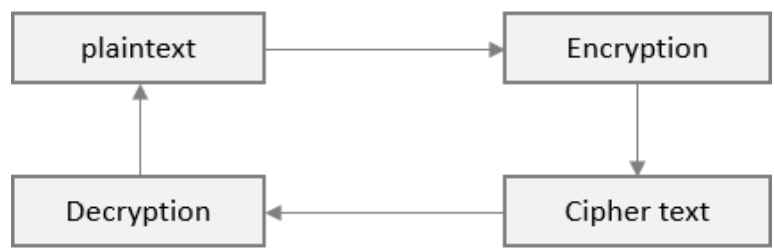

Figure 1. Cycle of Encryption

## 2. RELATED WORK

Cryptography involves converting standard text into unreadable form through encryption and reversing the process through decryption [15, 16]. The conventional encryption model follows four steps: sending plain text, converting it into cipher text using a key and algorithm, transmitting the cipher text, and converting it back to the original message at the receiver's end using the same algorithm [17].

Five main goals must be considered to maintain system secrecy [18, 19, and 20].
1. Authentication: Verify sender and receiver identities before message exchange.
2. Security/Privacy: Ensure only authenticated users can access the message.
3. Integrity: Prevent unauthorized alterations to the message during transmission.
4. Non-repudiation: Provide evidence of sender's authenticity to prevent false denial.
5. Service Reliability: Ensure system availability and protect against attacks for uninterrupted user experience.

Depending on the type of security keys used, cryptography is divided into two broad categories: symmetric and asymmetric encryption [21]. For encryption and decryption in symmetric encryption, also known as "private key," a similar key is used. Asymmetric encryption, also known as public-key encryption, employs two keys for encryption and decryption [22]. While symmetric encryption requires the sender and receiver to agree on the same secret key, asymmetric encryption uses two keys: a public key known to the public for encryption and a private key known to the client for decryption [23].

It also covered various modes of encryption and decryption [24, 25]:
1. Electronic Code Book (ECB): In this mode, data is divided into 64-bit blocks and encrypted individually. Each block operates independently, so transmission errors affect only the affected block. Since only the basic DES algorithm is used, ECB is the least secure mode.
2. Cipher Block Chaining (CBC): In CBC mode, each block of encrypted cipher text from ECB is XORed with the next block of plaintext. This creates dependencies between blocks, requiring knowledge of previous blocks to decrypt a specific block.
3. Cipher Feedback (CFB): CFB mode allows encryption of plaintext blocks smaller than 64 bits. It uses a 64-bit shift register as input for DES. This mode offers security but is slower than ECB due to increased complexity.
4. Output Feedback (OFB): OFB mode is like CFB, but the DES cipher text output is fed back into the shift register instead of the final cipher text.

## 2.1. AES implementation on Android for Message

Encryption algorithms play a crucial role in ensuring secure data transmission and protecting privacy [26]. With the increasing importance of cell phone security and the vulnerability of the SMS industry to cyber-attacks, message encryption has become a vital task [27]. Encryption has been historically used by militaries and governments for confidential communication, and it is now commonly employed in civilian systems to safeguard data [28, 29]. The Advanced Encryption Standard (AES) is a widely used encryption algorithm due to its efficient performance on various processors [30, 31].

AES offers block ciphers with key sizes of 128, 192, or 256 bits. The encryption process involves multiple rounds that convert plain text into cipher text. Each round's output becomes the input for the next round until the final round produces the cipher text. AES incorporates operations such as sub bytes, row shifting, column mixing, and the addition of a round key. With a 128-bit cipher key, breaking the encryption would require examining 128 possibilities, while the absence of fixed patterns enhances security [32, 33].

The application implementing AES encryption is compact (50 kilobytes) and compatible with Android phones. It meets speed requirements, providing a seamless user experience. The user interface is user-friendly, facilitating navigation. The program can authenticate message senders, ensuring access control. It also detects message corruption or modification during transmission [34]. Sensitive information within messages remains secure even if an adversary gains access to the device. The application prioritizes data security against brute force and pattern attacks. Additionally, it ensures secure end-to-end data transfer without corrupt message segments [35].

## 2.2. Security Analysis of AES Algorithm

The researchers examined various encryption algorithms, including AES, DES, and RSA, categorizing them into symmetric and asymmetric types. Symmetric encryption employs a single key for both encryption and decryption, while asymmetric encryption uses a public key and a private key to address the challenge of key distribution [36]. Asymmetric encryption, although slower due to increased computational demands, encrypts data with the public key and decrypts it with the private key. DES adopts 64-bit keys, while AES supports 128, 192, or 256-bit keys. The study detailed DES, a widely used method superseded by the faster AES algorithm incorporating sub-bytes, round keys, shift rows, and mix columns [37]. Comparative analysis encompassed 18 factors, indicating that AES had the shortest encryption time, outperforming RSA, and DES [38].

A cryptographic algorithm is vital for information security, encrypting text. Two types of algorithms exist: symmetric (private) key cryptography and asymmetric (public) key cryptography. Private key encryption employs a single key for both encryption and decryption, requiring key distribution before data transmission [39]. Weak keys undermine security, facilitating decryption. Key size determines strength, with larger keys being less susceptible to brute-force attacks. Symmetric algorithms like DES (56-bit key) and AES (128, 192, and 256-bit keys) are commonly used [40]. Public-key encryption employs private and public keys, solving the key distribution problem. The private key decrypts, while the public key encrypts. The public key is widely known, while the private key remains exclusive to the user, eliminating the need for prior key distribution [41]. Asymmetric key encryption relies on computationally demanding mathematical functions suitable for small mobile devices. In our Android application, we utilized the AES algorithm, an asymmetric encryption algorithm, for message encryption [42].

## 3. METHODOLOGY

We have proposed a set of protocols for the encryption and decryption of data to ensure secure transmission of information between a source and destination. The protocols aim to protect the confidentiality and integrity of the data during transit, mitigating the risk of unauthorized access and tampering. Our proposed solution incorporates state-of-the-art Advance Encrypted Standard (AES) algorithms to support robust data protection. The encryption process transforms the original data into an unintelligible form using AES and keys. This ensures that even if the transmitted data is intercepted, it remains unreadable to unauthorized parties. Upon reaching the destination, the encrypted data is decrypted using the corresponding decryption protocols and keys. This process reverts the data back to its original form, enabling the intended recipient to access and utilize the information securely. To further enhance security, our protocols employ key management mechanisms to securely generate, distribute, and store encryption keys. These keys are unique to each transmission session and are designed to be resistant to various cryptographic attacks. Overall, our proposed protocols offer a comprehensive solution for secure data transmission, safeguarding sensitive information against unauthorized access, interception, and tampering. By implementing these protocols, organizations can establish a secure communication channel between source and destination, protecting the confidentiality and integrity of their valuable data.

Further, our article is distributed in the following four sections 1) Ontological based Semantic Annotation, 2) Personalized way for Semantic Annotation Recommendation 3) Socio-Technical way for Semantic Annotation Sharing, and 4) Recommendation Representation Through Knowledge Graph where we have briefly explained each layer and approach of a proposed methodology.

### 3.1. Proposed Encryption Model

When the app is launched, it directs the user to the main page where they can send an SMS. Upon selecting this option, the app prompts the user for a mobile number and message. To send the message, the user simply presses the send button. At this point, the algorithm generates a random 256-bit key and takes the text and key as input. The generated key is used to encrypt the messages using a specific algorithm. The plain text of the SMS is then converted into cipher text through encryption. Once the SMS is successfully sent to the intended destination, a dialog box confirms the delivery, marking the end of the process (Figure 2).

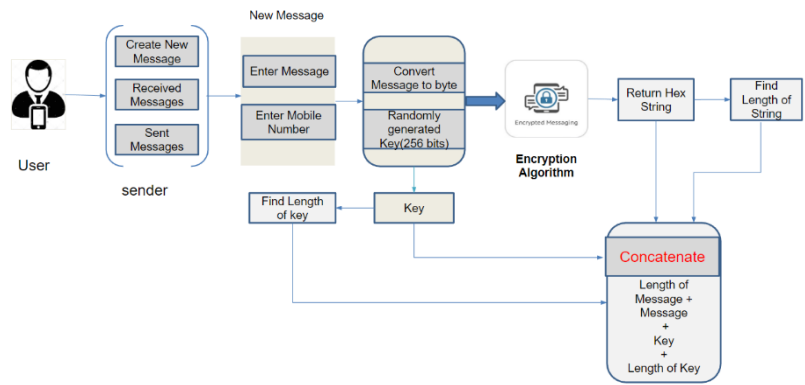

Figure 2. Use Case Diagram for sender-side encryption.

During this step (Figure 2), a 256-bit random key is generated, and the message is converted to a byte format. The encoded key and byte message are then passed to the algorithm, which returns an encrypted string. This encrypted string is subsequently typecast to a hexadecimal string. To maintain integrity, the key is converted to Base64 encoding, ensuring that even minor changes, such as spaces or commas, are preserved. Any alteration in the key would result in decryption exceptions. The variables, including the message length, message content, key, and key length, are concatenated into a single variable sent as a hexadecimal string.

### 3.2. Proposed Decryption Model

Upon receiving a message, it is initially stored in the built-in SQLite database of the mobile device. The application retrieves all the messages from the database and performs a specific operation to retrieve both the key and the encrypted message. Once the key and cipher text are obtained, the message is passed to the algorithm along with the necessary data. The algorithm decrypts the message, converting it back to plaintext, which is then displayed in the text field for the user to read. This marks the end of the process (Figure 3).

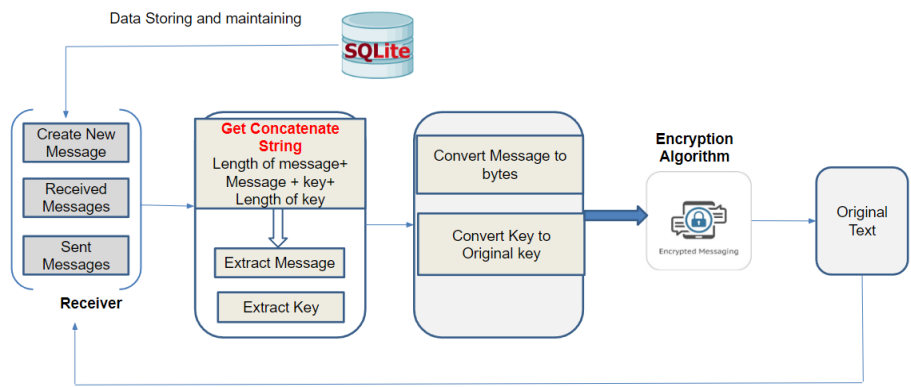

Figure 3. Use Case Diagram for message and key retrieval.

When a message is received, we extract the key and encrypted message separately. By determining the lengths of the key and message, we utilize mathematical operations to extract them from a combined hexadecimal string. Upon receiving the message, we decode it into bytes, like Base64 bytes. The key is decoded into the Secret key, which is essential for the algorithm.

To prevent specific exceptions like padding or incomplete blocks, we ensure that the key remains unchanged during encoding and decoding by using Base64. By having the key and message, we proceed to decrypt the cipher text, resulting in the algorithm returning the plain text (Figure 3).

### 3.3. Key Exchange Process

AES is a symmetric key algorithm, which means that it uses the same key for encryption and decryption. In AES, key exchange is a difficult task. In this application, we generate a key and use it to encrypt a message, using the same key that we send with the message. This technique differs on the sender and receiving sides, as explained further in Figure 4.

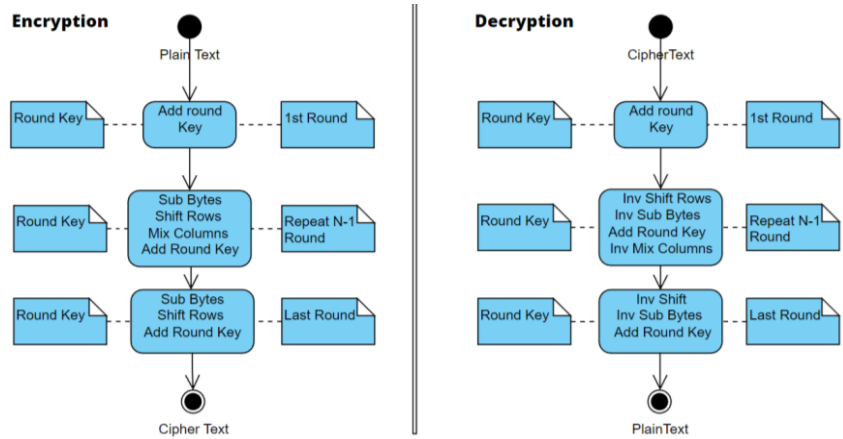

Figure 4. State Machine Diagram for AES Encryption & Decryption process

### 4. RESULT AND DISCUSSION

#### 4.1. Sender Side

To send SMS messages through the "Secure SMS App," users must set it as the default application upon launching. A prompt will appear, asking users to make it the default and enable encrypted messaging by clicking "Yes" (Figure5 a). The app also requires necessary permissions, such as storing, sending, and receiving SMS messages, for secure functionality (Figure 5 b).

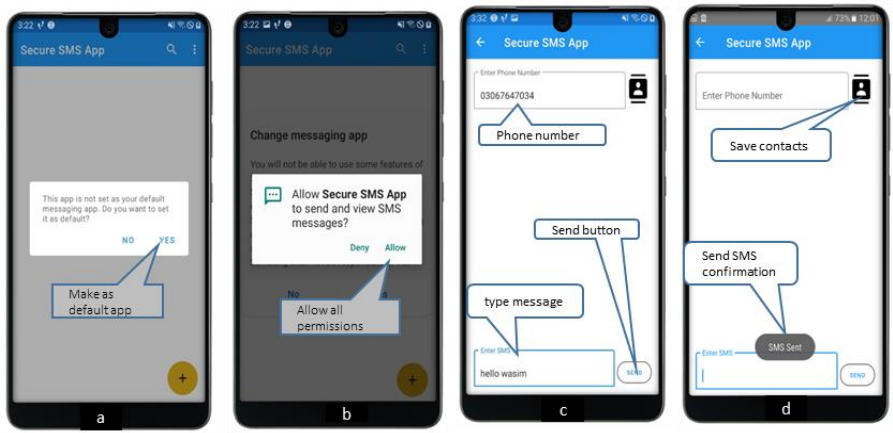

Figure 5. (a) Make as default app, (b) All Permissions, (c) Sending new SMS, (d) SMS sent

The user interface of the Secure SMS App is user-friendly. To send a message, users enter the message content and the sender's number in the app. Upon clicking the "send" button, the message is encrypted before being sent to the destination (Figure 5 c & d). A Toast message confirming the successful sending of the message is displayed, without revealing the encrypted content. The encrypted message can be viewed on the console (Figure 6).

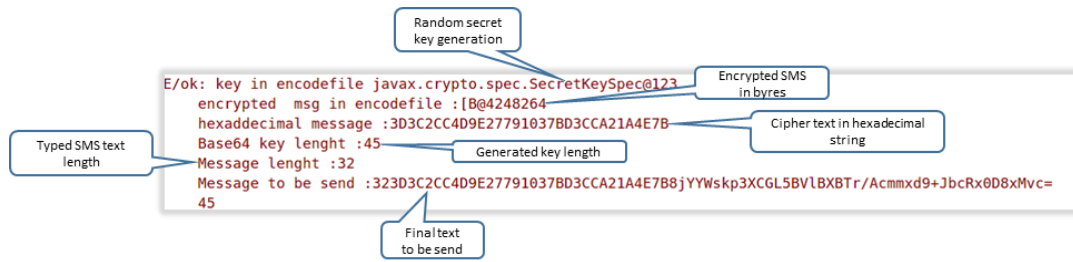

Figure 6. Result of Figure 5 (c & d) from the android studio console

### 4.2. Receiver End

When you receive messages, the application will retrieve a key and text from a combined hexadecimal string. The algorithm will process the key and message and return plain text. If you click on the message, you will see the original message without having to perform any additional actions (Figure 7).

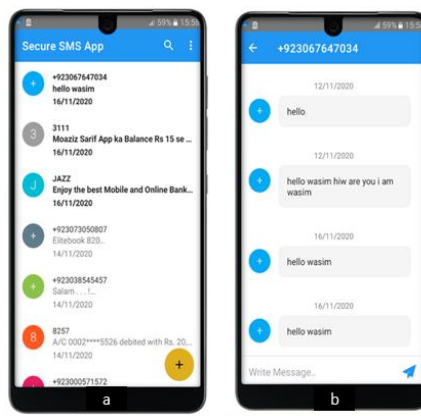

Figure 7. (a) All SMS Detail, (b) individual SMS Detail

### 4.3. Static and Dynamic testing using MobSF

To assess the security and protocol of our app, we used the popular Mobile Security Framework (MobSF). We tested both static and dynamic apps during testing. MobSF calculates our app's security score. Our paper, as we all know, focuses on SMS security rather than app security; however, app security is equally important, and we will cover it in a future edition of our app. We convert our mobile traffic to a PC and send SMS messages during testing. The MobSF tests each component of the app. Figures 8 and 9 depict the result.

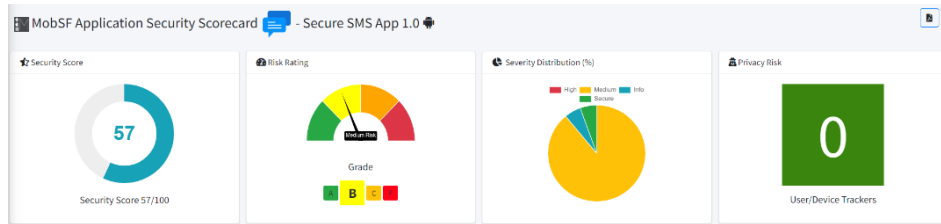

Figure 8. MobSF Security check result

Because the app uses various types of permissions, such as accessing contacts, storage, sending SMS, receiving SMS, and broadcast receiver, the overall security score is 57, which lowers the security score of a mobile app. The focus of this paper is on protecting SMS from call center staff and intruders during transmission. Furthermore, MobSF has a privacy risk score of 0 and a risk grade of B, indicating a medium risk level. As shown in FIGURE 9, MobSF also recommends more medium-risk content related to our app. It suggests that we reduce the application rate (Figure 9).

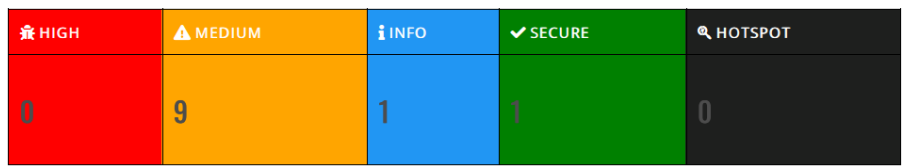

Figure 9. Final finding of MobSF

The investigation is centred on SMS security, and the communication protocol must be tested. MobSF is used to test the TLS/SSL protocol, and the results are shown in figure 10. During testing, we went through each screen of the app and sent SMS messages to get an exact result. All the Check Marks are green, indicating that the test was successful, and no vulnerabilities were found (Figure 10).

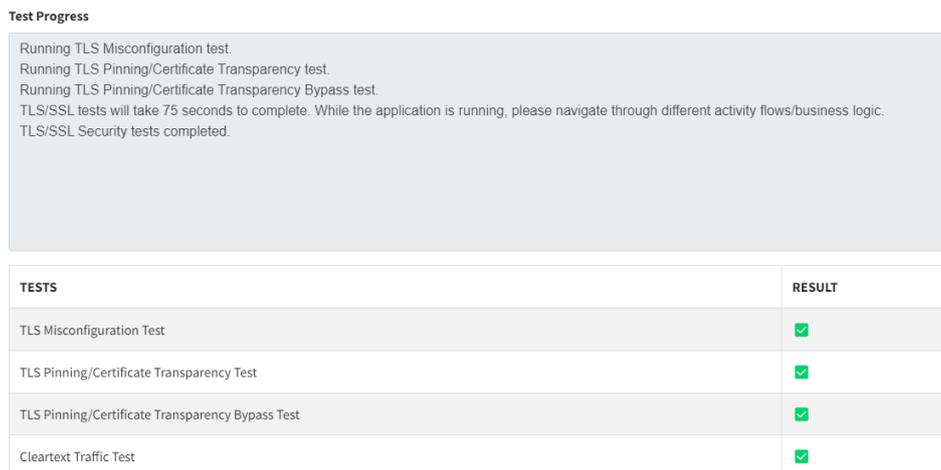

Figure 10. TLS/SSL Network Protocol

### 4.4. Manual Testing and Application Analysis

The app has been tested on a Samsung J7 running Marshmallow version 6. The app is installed on two devices, and SMS are sent in real time over the mobile network. The same messages were also tested on the phone's built-in messaging app, and there was a significant difference. The received SMS were displayed in encrypted form in the phone's built-in app, while the same SMS were displayed in plain text in the secure app, which is automatically decrypted by the app, as shown in the figures above. Table 3 displays the results for SMS sent, while Table 4 displays the results for SMS received. Several SMS messages have been sent to assist with the app's encryption and decryption. These tables show the results of the app's manual testing. Table 5 and 6 presented application storage and cache.

Table 1. A test case for sending an encrypted message.

| S. No | Encoded Key | Key to Base64 | Input SMS | Encrypted SMS in |
|---|---|---|---|---|
| 1 | [B@4248264 | 8jYYWskp3XCGL5BVlBXBTr/Ammxd9+JbcRx0D8xMvc= | hello wasim | 3D3C2CC4D9E27791037BD3CCA21A4E7B |
| 2 | [B@847e074 | 7jSWskp3XDHL5BVlBXBTr/Annxd9+JbcRx0D9YMvc= | Hello world | B54D78oAo5F663AAD8FC6E6FFCE3DCE2 |

Table 2. A test case for received encrypted message.

| Sr. No | Encoded Key in Bytes | Encrypted Message | Decrypted output / original text |
|---|---|---|---|
| 1 | [B@4248264 | 3D3C2CC4D9E27791037BD3CCA21A4E7B | Hello wasim |
| 2 | [B@847e074 | B54D78oAo5F663AAD8FC6E6FFCE3DCE2 | Hello world |

## CONCLUSIONS

SMS is the most widely used and important mode of communication. This data may contain sensitive and vital information that must be protected, which encryption can provide. We researched cryptographic algorithms for this purpose. Although asymmetric algorithms require two independent keys to encrypt and decrypt, we discovered that they use complex mathematical functions that are inefficient for small mobile devices. As a result, we employ a symmetric algorithm for encryption. Furthermore, among symmetric algorithms, AES is the most efficient and resistant to brute-force attacks. As a result, we created an Android app that assists the sender in encrypting the information using a key before sending it to the receiver, who can decrypt the message using the same key. It will be impossible to decrypt the messages while they are being transmitted. The SMSC Center staff and the attacker will not be able to decrypt the information during transmission because it will be scrambled.

## ACKNOWLEDGEMENTS

This research was supported by "Regional Innovation Strategy (RIS)" through the National Research Foundation of Korea (NRF) funded by the Ministry of Education (MOE) (2022RIS-005).